\documentclass[sigconf,10pt,screen,nonacm]{acmart}

\makeatletter
\renewcommand\@formatdoi[1]{\ignorespaces}
\makeatother

\AtBeginDocument{%
  \providecommand\BibTeX{{%
    \normalfont B\kern-0.5em{\scshape i\kern-0.25em b}\kern-0.8em\TeX}}}

\usepackage{titletoc}
\usepackage{url}
\usepackage{xcolor}
\usepackage{hyperref}

\renewcommand\footnotetextcopyrightpermission[1]{} 
\begin{document}

\title{SafetyOps}



\author{Umair Siddique}
\affiliation{\institution{Bosch Research and Technology Center (RTC)\\
    Sunnyvale, California,  USA}}
\email{umair.siddique@us.bosch.com}
\renewcommand{\shortauthors}{U. Siddique}


\begin{abstract}
Safety assurance is a paramount factor in the large-scale deployment of various autonomous systems (e.g., self-driving vehicles\footnote{The paper is specific to autonomous vehicles. However, the position taken in this paper may be applicable to other autonomous systems.}). However, the execution of safety engineering practices and processes have been challenged by an increasing complexity of modern safety-critical systems. This attribute has become more critical for autonomous systems that involve artificial intelligence (AI) and data-driven techniques along with the complex interactions of the physical world and digital computing platforms. In this position paper, we highlight some challenges of applying current safety processes to modern autonomous systems. Then, we introduce the concept of \textbf{SafetyOps} -- a set of practices, which combines DevOps, TestOps, DataOps, and MLOps to provide an efficient, continuous and traceable system safety lifecycle. We believe that SafetyOps can play a significant role in scalable integration and adaptation of safety engineering into various industries relying on AI and data. 
\end{abstract}




\maketitle
\pagestyle{empty}

\section{Introduction}

In the last decade, artificial intelligence (AI) and automation have become an integral part of our life.  The range of applications varies from chatbots,  smart homes to robotic surgery, and automated cars.  Any safety-critical (e.g., automated cars, surgical robots, and avionics systems) applications of such technology require a rigorous safety assurance before large-scale deployment.  Consequently, manufacturers and developers usually follow a safety engineering lifecycle per international standards such as ISO 26262 \cite{ISO26262}, ISO/PAS 21448 \cite{ISO/PAS21448}, IEC 61508 \cite{IEC61508},  and EN 50128 \cite{EN50128} depending upon the application domain. The ultimate goal is to ensure the functional safety (FuSa) and safety-of-the-intended  functionality (SOTIF), which deal with the absence of unreasonable risks due to hazards caused by malfunctioning behavior of software/hardware components, algorithmic insufficiencies, and foreseeable misuse of the underlying technology.

In theory, safety assurance is a continuous and iterative process that systematically provides confidence that an underlying product meets the safety requirements. However, manual execution of the safety processes makes it challenging to maintain the continuity of safety assurance throughout the product lifecycle. The functional safety requirements and associated tests for autonomous vehicles heavily depend on (training) data and operational design domain (ODD) \cite{Czarnecki2018,koopan2016}.  However, dynamically linking a given ODD and data to safety requirements, system architecture, safety artifacts, and test results is a nontrivial task. Moreover, the development of autonomous systems  can benefit from both \emph{requirement specification} at design time and \emph{requirement mining} from the real data. Though integrating feedback loops in requirement management tools is a challenging task and involves the development of custom plugins using representational state transfer (REST) application programming interfaces (APIs).

In this paper, we highlight that effective, efficient, and continuous safety assurance of complex systems call for significant updates in the safety processes, analysis techniques, and verification methodologies. We propose SafetyOps as an emerging area that can combine the power of various automation frameworks (e.g., DevOps, TestOps, DataOps) to provide continuous delivery of safety assurance cases. Moreover,  SafetyOps can reduce the gap between safety, software, hardware, and test engineers.

The organization of the paper is as follows: Section \ref{challenges} enumerates the current and upcoming challenges related to safety assurance. We provide the principles of SafetyOps in Section \ref{found}. Finally, Section \ref{con} concludes the paper.

\section{Challenges} \label{challenges}

\begin{itemize}
\item \textbf{Gap in safety engineering and software development frameworks:}
Safety standards provide some guidelines for preparing safety documents (e.g., safety concept) and performing various safety analyses (e.g., hazard and risk analysis (HARA) and fault tree analysis (FTA)). However, the execution of these processes is quite different from modern software workflows. For example, visualizing and reviewing a large FMEA or FTA is not as structured as Git code reviews. Indeed, safety processes are largely isolated from the design phase and connected only by manual processes, Excel spreadsheets, and human middleware. Traditionally, we would argue that safety activities have to be performed independent from the design and development teams.
These seemingly contradicting requirements induce a massive challenge for developing autonomous systems --
to perform a holistic SOTIF analysis of robotic algorithms intrinsically requires the involvement and engagement of both developers and safety engineers.


\item \textbf{Detachment of various safety analysis methods:}
In practice, various safety analyses (e.g., FMEA and  FTA) documents for a given system are produced by the safety team. However,  consistency checking among these analysis documents is usually done manually. For autonomous systems, these analysis documents need to be linked dynamically.
Usually,  FTA and FMEA are manually performed at design time to gain more insights into the system.  However, current AI-enabled autonomous systems collect plenty of data during the trial phase and provide the opportunity to learn a system fault tree from the data \cite{Lift}. Similarly, quantitative FMEA requires to mine the frequency of failure modes from the real data.
The traditional FMEA and FTA  methods have some limitations to capture the complicated details of the modern system, e.g.,  \cite{Rick2019} proposed a specialized FMEA for perceptual components in automated driving. Similarly, various ODD analysis methods (e.g., \cite{koopman2019autonomous}) are not readily supported in existing safety tools.
We believe that the effectiveness of failure analysis methods can be improved by considering these attributes  in a unified framework.

\item \textbf{Continuous and traceable delivery of safety cases:}

A safety case usually depends on the system design, environmental assumptions, system configuration, identified hazards,  associated risks, and corresponding safety measures.
However, in the case of AI-enabled autonomous systems deployed in ever-changing public environments (e.g. self-driving vehicles), these attributes can change more often than for some more restricted systems (e.g., factory automation robots). This situation would benefit from the development of machine-readable, hierarchical \cite{HierarichalSC}, and dynamic \cite{DynamicSC} safety cases. Indeed, this requires a well-connected framework for system design, data mining, safety, and testing artifacts.

\item \textbf{From simulation to real experiments and back:}
The certification and safety assurance would require autonomous systems to operate over real-world environment dictated by the chosen ODD. However, infinite state-space due to an open context makes it impossible to test all corner cases in a real-world setting. There is a growing consensus in the industry that simulation \cite{simulation} (either using fenced-road-network or using software) need to play a role along with the real-world testing. It leads to an absolute dependence on creating a scenario library \cite{Scenario} to drive the simulation-based testing. The scenarios can be hand-crafted or learned from the real data -- furthermore, randomization can be used to verify robustness related properties of the system. Here, a few aspects are critical: 1) validation of such scenarios with respect to an ODD, 2) establishing a link between scenarios and other functional safety and SOTIF artifacts, and 3) quantification of the domain coverage.

\end{itemize}

\section{Principles of SafetyOps} \label{found}
In the last few years, DevOps  \cite{devops} has revolutionized the software development and services industry \cite{devops-report}. Indeed, DevOps has now become a part of almost every industry ranging from web technologies and banking sectors to automotive and avionics. Following this development, similar approaches have emerged in other fields, such as data engineering, machine learning, and system testing.   In the following, we define them\footnote{Note that this list is not exhaustive. Other fields where \emph{Ops}-ing has emerged include PrivacyOps, SecurityOps, SysOps, and ModelOps.}:
\begin{figure*}[hbt]
	\vspace{-0.17in}
  \includegraphics[width=1.0\textwidth]{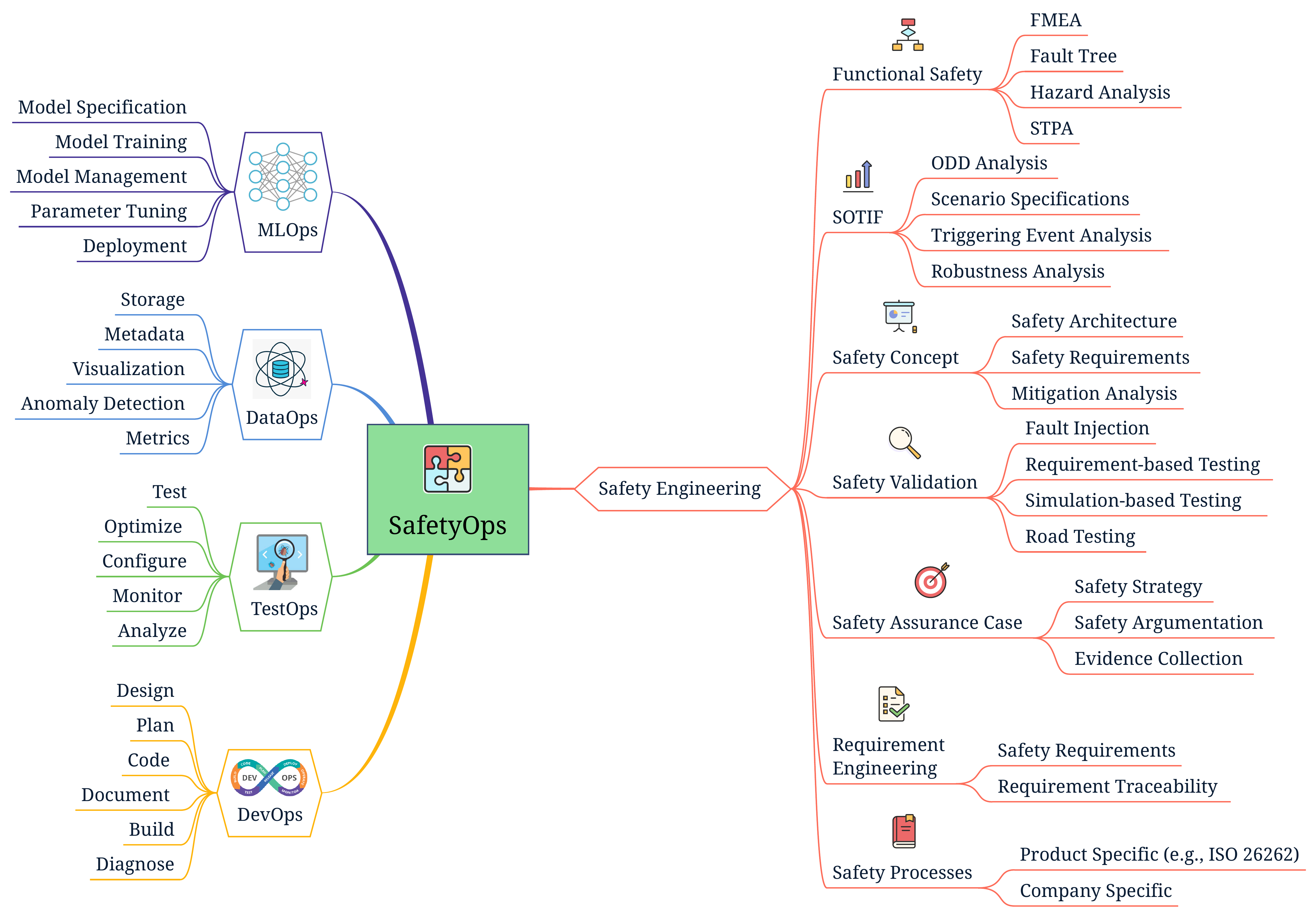}
  \caption[Caption for LOF]{An Overview of SafetyOps: Note that the branches (on the right) representing various safety analysis techniques are not exhaustive. The primary purpose is to show that these techniques require a dynamic link with the activities of DevOps, TestOps, DataOps, and MLOps.}
   \label{Fig:safetyops}
\end{figure*}

\begin{itemize}

\item \textbf{DevOps} is a set of practices that combines software development (Dev) and information-technology operations (Ops), which aims to shorten the systems development life cycle and provide continuous delivery with high software quality.

\item \textbf{TestOps} is a combination of Test and Operations (Ops) -- a more evolved version of traditional testing to assure the quality of services.

\item \textbf{DataOps} is an automated, process-oriented methodology, used by data engineers, to improve the quality and reduce the cycle time of data analytics.  DataOps applies to the entire data lifecycle, from data preparation to reporting.

\item \textbf{MLOps}  typically refers to the collaboration between data/machine learning scientists and operations engineers to manage the lifecycle of ML within an organization.

\end{itemize}

It is important to note that the frameworks mentioned above are closed-loop, which allows for fast feedback and continuous improvement based on the learning/data produced in each iteration.  Moreover, feedback loops ensure quality and measurable performance throughout the system lifecycle. A significant source of their widespread acceptance in industry is the abundance of open-source contributions. One would wish to adopt these frameworks directly for safety engineering to cope with the challenges posed by the modern autonomous systems (mentioned in Section \ref{challenges}). However, we believe that safety has its own unique needs that require a dedicated framework (we call it as SafetyOps). \\

\fbox{\begin{minipage}[t]{0.45\textwidth}
\textbf{SafetyOps} combines DevOps, TestOps, DataOps and MLOps with \emph{safety engineering} to provide a fast, efficient, continuous and traceable system safety lifecycle.
\end{minipage}} \\

An overview of SafetyOps is shown in Figure \ref{Fig:safetyops}. The main idea is to build a framework that seamlessly links safety engineering activities (represented by branches in Figure \ref{Fig:safetyops}) to essential components for autonomous systems development (i.e., DevOps, TestOps, DataOps, and MLOps). Indeed, this integration greatly improves and accelerates the safety assurance activities.  Achieving a SafetyOps environment requires changes in processes, tools, and culture. We believe that the right tools and processes can provide measurable improvements in the following critical areas.

\begin{itemize}

\item \textbf{Reduction in retrospective activities:} Due to a gap between safety engineers and other teams, some activities are usually performed in a retrospective fashion. With the SafetyOps framework in place, these activities can be performed continuously during the development.

\item \textbf{More acceptance of safety in development teams:} Modernization of safety tools and processes provided by SafetyOps can help in the acceptance of safety activities in development teams. For example, the concept of doc-as code\footnote{\url{https://www.writethedocs.org/guide/docs-as-code/}} can significantly improve the development and traceability of safety requirements.

\item \textbf{Continuous development of safety cases:}
Continuous integration of safety into the development process provided by SafetyOps allows realizing a safety case as a live artifact.
Indeed, this reduces the manual efforts of evidence collection and manual checking of changes/updates in any design, testing, or safety artifact. Moreover, automation of these activities can also help in measuring the impact of specific components/activities on the safety case and hence significantly elevate the safety of the released product.

\end{itemize}

Note that the core idea of SafetyOps is to promote  the automation (as in DevOps, TestOps, DataOps and MLOps) of safety processes and activities in accordance with corresponding standards (e.g., \cite{ISO26262}, ISO/PAS 21448 \cite{ISO/PAS21448}).
The links in Figure \ref{Fig:safetyops} can be realized in various ways\footnote{The details about the implementation of these links is beyond the scope of this paper.} depending upon the organizational setup (i.e., data-pipeline, development framework, requirement management and verification \& validation strategy).

In the following, we describe the main principles of the SafetyOps framework:

\begin{itemize}
	\item \textbf{Common data exchange and data management:}
The core of SafetyOps workflow is functionality for capturing and exchanging data between different tools (e.g., different FMEA, FTA, and safety case development tools). It also includes scenario benchmarks and test results. The ongoing developments in IEEE P2851\footnote{\url{https://sagroups.ieee.org/2851/}} can support this aspect by providing an exchangeable format for safety analysis and safety verification activities.

\item  \textbf{Open APIs and qualifiable open-source tools:}
Common data exchange and data management start with application programming interfaces (APIs) that enable the exchange of data between multiple software components, both vendor-supplied and in-house developed. For example, APIs can help in using driving scenarios in various simulation engines. Similarly, APIs can be used to find the correspondence between failure modes in FMEA and basic events in FTA  with the recorded data.

\item \textbf{State-of-the art visualization techniques and integrated analytics:} 
The modern interactive software and data visualization techniques can improve the development and outlook of safety artifacts. Such techniques indeed provide the basis to build navigable safety cases, fault trees, and hierarchical visualization of hazard analysis. Moreover, integrated analytics is essential to continuously monitor critical safety metrics, e.g., ODD coverage per test run, frequency of a particular failure mode in the recorded data, and average reaction time for a specific situation during the on-road testing.

\item\textbf{Inclusiveness of data-driven safety:}  Finally, the SafetyOps supports the integration of data-driven safety aspects.  For example, ideas around mining scenarios from the data, learning fault trees, and learning autonomous agent behavior invariance over a given ODD.
\end{itemize}

\section{Conclusion} \label{con}

In this position paper, we argue that safety engineering needs a framework like SafetyOps to handle the complexity of modern autonomous systems in the context of ever shorter product life cycles. SafetyOps can play a role in solving some crucial challenges related to the development of autonomous vehicles. An ultimate objective of SafetyOps is to automate the manual safety processes, provide continuous traceability between safety artifacts and main activities of the system development pipeline (data engineering, machine-learning, and system integration and testing). Finally, SafetyOps can also help in distributed development by utilizing the standardized interfaces for safety artifacts.



\bibliographystyle{ACM-Reference-Format}
\bibliography{biblio}

\end{document}